\documentclass[12pt]{article}
\usepackage{a4wide}
\usepackage{amssymb}
\begin{document}
{\renewcommand{\thefootnote}{\fnsymbol{footnote}}
\begin{center}
{\LARGE  Time in quantum cosmology}\\
\vspace{1.5em}
Martin Bojowald\footnote{e-mail address: {\tt bojowald@gravity.psu.edu}}
and Theodore Halnon\footnote{e-mail address: {\tt trh5241@psu.edu}}
\\
\vspace{0.5em}
Institute for Gravitation and the Cosmos,\\
The Pennsylvania State
University,\\
104 Davey Lab, University Park, PA 16802, USA\\
\vspace{1.5em}
\end{center}
}

\setcounter{footnote}{0}

\begin{abstract}
  A cosmological model with two global internal times shows that time
  reparameterization invariance, and therefore covariance, is not guaranteed
  by deparameterization. In particular, it is impossible to derive proper-time
  effective equations from a single deparameterized model if quantum
  corrections from fluctuations and higher moments are included. The framework
  of effective constraints shows how proper-time evolution can consistently be
  defined in quantum cosmological systems, such that it is time
  reparameterization invariant when compared with other choices of coordinate
  time. At the same time, it allows transformations of moment corrections in
  different deparameterizations of the same model, indicating partial time
  reparameterization of internal-time evolution. However, in addition to
  corrections from moments such as quantum fluctuations, also factor ordering
  corrections may appear. The latter generically break covariance in
  internal-time formulations. Fluctuation effects in quantum cosmology are
  therefore problematic, in particular if derivations are made with a single
  choice of internal time or a fixed physical Hilbert space.
\end{abstract}

\section{Introduction}

Deparameterization has become a popular method to circumvent the problem of
time in canonical quantum gravity. Since coordinate time is observer-dependent
and does not have a corresponding operator after quantization, one instead
selects one of the phase-space degrees of freedom as a measure of change for
other variables \cite{GenHamDyn1,BergmannTime}. Popular examples of internal
times are a free massless scalar field or a variable that quantifies dust.

These variables are turned into operators when the theory is quantized and
therefore appear in the state equations. They are of such a form that
constraint equations can be rewritten as familiar evolution equations, for
instance of Schr\"odinger or Klein--Gordon type. However, as part of the
general problem of time \cite{KucharTime,Isham:Time,AndersonTime} there is
some arbitrariness involved in the choice of a particular internal time. Just
as with coordinate time in classical general relativity or its cosmological
models, one would therefore like to show that the choice of internal time does
not affect predictions made from a quantum cosmological model. Only then can
the model and its underlying theory be considered covariant.

The question of covariance in internal-time formulations has rarely been
studied, but some results are available \cite{MultChoice,ClocksDyn}. In this
paper, we use semiclassical methods developed for effective constraints
\cite{EffCons,EffConsRel,EffTime,EffTimeLong,EffTimeCosmo} in order to study
this quation. We analyze an explicit model which permits two different choices
of internal time. At a semiclassical level, the methods of effective
constraints will be used to demonstrate covariance of moment corrections in
the two internal-time formulations. However, the introduction of a proper-time
parameter turns out to be a more complicated step than usually
appreciated. Such a parameter is important in order to relate evolution
equations to observer frames, and it is often used in quantum cosmology in
order to reformulate quantum evolution equations as effective or modified
Friedmann equations.

In usual treatments of this question, it seems to be assumed implicitly that
time reparameterization invariance is always guaranteed in homogeneous models
of quantum cosmology because they are subject to just one constraint, the
Hamiltonian constraint $C$ with spatially constant lapse function. A single
constraint always commutes with itself and therefore remains first class even
if it is modified by quantum effects or fully quantized. 

The last statement is correct, but it cannot always be applied to homogeneous
quantum cosmology. In a Dirac quantization of a homogeneous cosmological model
one replaces the classical constraint equation $C=0$ by an equation
$\hat{C}\psi=0$ for physical states $\psi$. Since solving the state equation
and constructing a suitable physical Hilbert space are complicated tasks, one
often takes a shortcut and computes an ``effective'' equation which can more
easily be analyzed, and which one expects to take the form of the classical
Friedmann equation plus quantum corrections. There are different procedures
for deriving such equations, but in some way they all make use of the
expectation value $\langle\hat{C}\rangle$ of the constraint operator in a
certain class of states. (The most systematic procedure of this type is the
canonical effective one already mentioned; see for instance
\cite{BounceSqueezed} for cosmological effective equations.)

The effective constraint equation $\langle\hat{C}\rangle=0$ then resembles the
Friedmann equation, as desired. But it does not imply that the state $\psi$
used in it is a physical state satisfying $\hat{C}\psi=0$. The quantum
constraint equation amounts to more than one independent expectation-value
equation, as systematically described in the formalism of effective
constraints. For instance, if $\hat{O}$ is some operator not equal to a number
times the identity, the equation $\langle\hat{O}\hat{C}\rangle=0$ is,
generically, independent of the equation $\langle\hat{C}\rangle=0$. The
premise in the tacit assumption that time reparameterization invariance is
always respected in homogeneous quantum cosmology because there is just a
single constraint is therefore violated. Making sure that time
reparameterization invariance, or more generally covariance, is still realized
after quantization, or under which conditions it can be broken, is then an
important task of quantum cosmology.

We will perform this task in the present paper for a specific model, and
confirm that covariance cannot be taken for granted in deparameterized
constructions. We then use the framework of effective constraints in order to
compare different deparameterizations within the same setting, which is made
possible by an analysis of the underlying gauge structure of quantum
constraints. This discussion will lead us to a general definition of
proper-time evolution in effective equations such that time reparameterization
invariance is realized in moment corrections. Our new definition leads to
proper-time evolution equations with moment corrections which are obtained
from those in deparameterized evolution by a change of gauge. Compared with
traditional derivations of proper-time evolution from deparameterized
evolution, however, the covariant formulation predicts different quantum
corrections for effective equations. A proper investigation of time
reparameterization invariance is therefore crucial for a reliable
determination of fluctuation corrections in quantum cosmological models.

In addition to moment corrections, different choices of internal or proper
time may give rise to different factor ordering corrections. In contrast to
moment corrections, these terms cannot be related by gauge transformations
because effective constraints and the gauge they generate are computed for a
fixed factor ordering. Factor ordering corrections therefore break covariance
of internal-time formulations. In our specific model, all three time choices
require different factor orderings of the constraint operator for real
evolution generators. Time reparameterization invariance is therefore broken
in internal-time quantum cosmology if all relevant corrections are taken into
account, a result which makes the outcome of \cite{MultChoice} more
specific. However, the new proper-time evolution introduced here is time
reparameterization invariant when compared with other choices of internal
time. This evolution is unique in realizing the same type of covariance as in
classical cosmological models, which is broken in internal-time evolution.

\section{The model}

Our cosmological model is isotropic, spatially flat, and has a cosmological
constant $\Lambda$ as well as a free, massless scalar field
$\tilde{\phi}$. Its classical description is therefore given by the Friedmann
equation
\begin{equation} \label{Friedmann}
 H^2= \frac{8\pi G}{3} \frac{\tilde{p}_{\tilde{\phi}}^2}{2a^6}+ \Lambda
\end{equation}
for the scale factor $a$ in $H=\dot{a}/a$ in terms of proper
time. 

We introduce the following canonical variables. (See \cite{ROPP} for a review
of quantum cosmology and of the notation used here.) The Hubble parameter $H$
is canonically conjugate to the ``volume'' 
\begin{equation}
 V:= \frac{a^3}{4\pi G}
\end{equation}
such that $\{H,V\}=1$. The scalar field $\tilde{\phi}$ is canonically
conjugate to the momentum $p_{\tilde{\phi}}$, such that
$\{\tilde{\phi},p_{\tilde{\phi}}\}=1$. The cosmological constant
$\Lambda$ is canonically conjugate to a variable which we call
$T$, such that $\{T,\Lambda\}=1$. 

The last statement may be unexpected. The cosmological constant is usually
treated as just that, a constant that appears in Einstein's equation much like
a fundamental constant such as $G$. However, it is mathematically consistent
to treat it as the momentum of a variable $T$ which does not appear in
the action or Hamiltonian constraint of the theory. The momentum
$\Lambda$ of any such quantity is conserved in time, and therefore
appears just as a constant in the field equations. We are not modifying the
dynamics by introducing this new canonical pair $(T,\Lambda)$,
nor are we trying to derive a mechanism for dark energy. We are merely using a
mathematically equivalent formulation of the usual theory, as will be clear
from the equations derived below. The new parameter $T$ then presents
to us a new option of a global internal time, which we can compare with the
more standard global internal time $\tilde{\phi}$.

We note that we do not intend $T$ to have physical meaning or to be
measurable. This might be taken as a disadvantage of the formulation, but it
is not that much different from the free scalar field $\tilde{\phi}$ for which
no physical explanation is known. Both fields are introduced primarily for the
purpose of serving as a global internal time. The variable $T$ in fact
has an advantage compared with $\tilde{\phi}$ because the energy density
associated with it is just the cosmological constant, for which there is
observational support. The energy density of a free scalar field, by contrast,
has not been observed. 

We have put tildes on the scalar symbols used so far. We now rescale these
quantities so as to remove most numerical factors from our equations, just for
the sake of convenience and in order not to distract from the important
terms. We introduce
\begin{equation}
 p_{\phi}:= \frac{p_{\tilde{\phi}}}{\sqrt{12\pi G}}
\end{equation}
and its canonical momentum $\phi$. It is straightforward to confirm that the
Friedmann equation (\ref{Friedmann}) is equivalent to the constraint equation
\begin{equation} \label{C}
 C=-VH^2+\frac{p_{\phi}^2}{V}+V\Lambda=0
\end{equation}
in these new variables. We have multiplied the terms in the Friedmann equation
with $V$ in order to have energies rather than energy densities. 

If we use this constraint to generate evolution equations with respect to
proper time, we should remember the factor of $(4\pi G)^{-1}$ in the
definition of $V$. The usual generator of proper-time evolution equations is
\begin{equation}
 \tilde{C}=\frac{3}{8\pi G} \frac{a^3}{V}C= \frac{3}{2}C\,.
\end{equation}
Our proper-time equations therefore differ by a factor of $3/2$ from the usual
ones, for instance
\begin{equation}
 \frac{{\rm d}V}{{\rm d}\tau}= \{V,C\}= 2VH
\end{equation}
which implies 
\begin{equation}
 H=\frac{1}{2V} \frac{{\rm d}V}{{\rm d}\tau}= \frac{3}{2a}\frac{{\rm d}a}{{\rm
     d}\tau} 
\end{equation}
with the promised factor of $3/2$ compared with the usual $H=\dot{a}/a$. For
completeness, we note the second classical evolution equation
\begin{equation}
 \frac{{\rm d}H}{{\rm d}\tau} = -H^2-\frac{p_{\phi}^2}{V^2}+\Lambda \approx
 -2(H^2-\Lambda) \approx -2 \frac{p_{\phi}^2}{V^2}
\end{equation}
with the last two weak equalities indicating that the constraint (\ref{C}) has
been used.

\section{Deparameterization}
\label{s:Deparam}

We deparameterize the model in two different ways, using  the global internal
times $\phi$ and $T$, respectively. We begin with the more familiar choice
$\phi$, solving $C=0$ for the momentum
\begin{equation}
 p_{\phi}(V,H,\Lambda) = -V\sqrt{H^2-\Lambda}\,.
\end{equation}

\subsection{Scalar time}

In this section, we quantize the model after deparameterization, so that there
is an operator $\hat{p}_{\phi}$ acting on a (physical) Hilbert space of wave
functions that do not depend on $\phi$, for instance $\psi(V,T)$. All we
assume about this operator for our semiclassical analysis is that it is Weyl
ordered. The methods of \cite{EffAc,Karpacz} then allow us to compute an
effective Hamiltonian by formally expanding the expectation value
\begin{eqnarray}
 H_{\phi}&:=&\langle p_{\phi}(\hat{V},\hat{H},\hat{\Lambda})\rangle = \langle
 p_{\phi}(V+(\hat{V}-V),H+(\hat{H}-H),\Lambda+(\hat{\Lambda}-\Lambda))\rangle\\
 &=& p_{\phi}(V,H,\Lambda)+\sum_{a_1,a_2,a_3=2}^{\infty} \frac{1}{a_1!a_2!a_3!}
 \frac{\partial^{a_1+a_2+a_3} p_{\phi}(V,H,\Lambda)}{\partial V^{a_1}\partial
   H^{a_2}\partial\Lambda^{a_3}} \Delta(V^{a_1}H^{a_2}\Lambda^{a_3})
\end{eqnarray}
in $\hat{V}-V$, $\hat{H}-H$ and $\hat{\Lambda}-\Lambda$. 

Although we use the same symbols $V$, $H$ and $\Lambda$ for our basic
variables, they now refer to expectation values of the corresponding
operators. In the expanded expression, in addition to expectation values, we
have the moments
\begin{equation}
 \Delta(O_1^{a_1}\cdots O_n^{a_n}) = \langle
 (\hat{O}_1-O_1)^{a_1}\cdots
 (\hat{O}_n-O_n)^{a_n}\rangle_{\rm symm}
\end{equation}
(with totally symmetric or Weyl ordering) as independent variables. For
instance, $\Delta(H^2)=(\Delta H)^2$ is the square of the $H$-fluctuation. If
the cosmological constant is just a constant, the quantum state is an
eigenstate of $\Lambda$, such that all moments including $\Lambda$
vanish. However, we keep these moments in our equations for full
generality. We will work exclusively with semiclassical approximations of the
order $\hbar$, which includes corrections linear in second-order moments or
terms with an explicit linear dependence on $\hbar$. We will ignore all
higher-order moments as well as products of second-order moments. The
elimination of higher-order terms will not always be indicated explicitly but
holds throughout the paper.  In our specific example, we have
\begin{eqnarray} \label{Hphi}
 H_{\phi} &=& -V\sqrt{H^2-\Lambda}- \frac{H}{\sqrt{H^2-\Lambda}} \Delta(VH)+
 \frac{1}{2} \frac{V\Lambda}{(H^2-\Lambda)^{3/2}} \Delta(H^2)\\
&&+
 \frac{1}{2\sqrt{H^2-\Lambda}} \Delta(V\Lambda)- \frac{1}{2}
   \frac{VH}{(H^2-\Lambda)^{3/2}} \Delta(H\Lambda)+ \frac{1}{8}
   \frac{V}{(H^2-\Lambda)^{3/2}} \Delta(\Lambda^2)\,. \nonumber
\end{eqnarray}

The commutator of operators induces a Poisson bracket on expectation values
and moments, seen as functions on the space of states. They can be derived
from the definition
\begin{equation} \label{AB}
 \{A,B\} = \frac{\langle[\hat{A},\hat{B}]\rangle}{i\hbar}
\end{equation}
and the Leibniz rule. In particular, the classical bracket $\{H,V\}=1$ still
holds true for the expectation values, and expectation values have zero
Poisson brackets with the moments. For Poisson brackets of two moments there
are general equations \cite{EffAc,HigherMoments}, but for small orders it is
usually more convenient to compute brackets directly from (\ref{AB}). For
instance, 
\begin{eqnarray}
 \{\Delta(H^2),\Delta(V^2)\} &=& 4\Delta(VH)\,,\\
 \{\Delta(H^2),\Delta(VH)\}&=& 2\Delta(H^2)\,,\\
 \{\Delta(V^2),\Delta(VH)\}&=&-2\Delta(V^2)\,.
\end{eqnarray}

These Poisson brackets give rise to the equations of motion
\begin{eqnarray}
 \frac{{\rm d}V}{{\rm d}\phi} &=& \{V,H_{\phi}\} =
 \frac{VH}{\sqrt{H^2-\Lambda}}- \frac{\Lambda}{(H^2-\Lambda)^{3/2}} \Delta(VH)
 +\frac{3}{2} \frac{VH\Lambda}{(H^2-\Lambda)^{3/2}} \Delta(H^2)\\
 &&+
 \frac{H}{2(H^2-\Lambda)^{3/2}} \Delta(V\Lambda)- \frac{1}{2}
 \frac{V(2H^2+\Lambda)}{(H^2-\Lambda)^{5/2}} \Delta(H\Lambda)+ \frac{3}{8}
 \frac{VH}{(H^2-\Lambda)^{5/2}} \Delta(\Lambda^2) \nonumber
\end{eqnarray}
and
\begin{equation}
 \frac{{\rm d}H}{{\rm d}\phi} = -\sqrt{H^2-\Lambda} + \frac{1}{2}
 \frac{\Lambda}{(H^2-\Lambda)^{3/2}} \Delta(H^2) -\frac{1}{2}
 \frac{H}{(H^2-\Lambda)^{3/2}} \Delta(H\Lambda)+ \frac{1}{8}
 \frac{1}{(H^2-\Lambda)^{3/2}} \Delta(\Lambda^2)\,,
\end{equation}
accompanied by equations of motion for the moments such as 
\begin{equation}
 \frac{{\rm d}\Delta(V^2)}{{\rm d}\phi} = 2\frac{H}{\sqrt{H^2-\Lambda}}
 \Delta(V^2)- 2\frac{V\Lambda}{(H^2-\Lambda)^{3/2}}\Delta(VH) +
 \frac{VH}{(H^2-\Lambda)^{3/2}} \Delta(V\Lambda)\,.
\end{equation}
Expectation values and moments are therefore dynamically coupled.

These equations can be compared with the classical Friedmann equation if we
transform them to proper time. The usual way to do so is by using the chain
rule after computing ${\rm d}\phi/{\rm d}\tau=\{\phi,C\}$. However, within the
deparameterized setting, we do not have a quantum-corrected expression for $C$
since we quantized $p_{\phi}$ after solving $C=0$. The introduction of proper
time in a deparameterized setting is therefore ambiguous. We will present two
different alternatives in this section, none of which will turn out to be
consistent by our general analysis in the next section.

The term in the constraint relevant for $\{\phi,C\}$ is $p_{\phi}^2/V$, while
the other two terms have zero Poisson brackets with $\phi$. We tentatively
introduce quantum corrections of this term by using the same methods that gave
us the quantum corrected $p_{\phi}(V,H,\Lambda)$. The new term is then
\begin{equation}
 C_{\phi}:= \frac{p_{\phi}^2}{V} - 2\frac{p_{\phi}}{V^2}
 \Delta(Vp_{\phi})+ \frac{p_{\phi}^2}{V^3} \Delta(V^2) +
 \frac{1}{V}\Delta(p_{\phi}^2)  \,,
\end{equation}
leading to 
\begin{eqnarray}
 \frac{{\rm d}\phi}{{\rm d}\tau} &=& \{\phi,-C_{\phi}\} = -2\frac{p_{\phi}}{V} +
 \frac{2}{V^2} \Delta(Vp_{\phi})-\frac{2p_{\phi}}{V^3} \Delta(V^2)\\
&=& 2\sqrt{H^2-\Lambda}+ 2\frac{\sqrt{H^2-\Lambda}}{V^2}\Delta(V^2)+ 
 \frac{2H}{V\sqrt{H^2-\Lambda}}\Delta(VH)-
 \frac{\Lambda}{(H^2-\Lambda)^{3/2}} \Delta(H^2)\\
 && - \frac{1}{V\sqrt{H^2-\Lambda}} \Delta(V\Lambda)+
 \frac{H}{(H^2-\Lambda)^{3/2}} \Delta(H\Lambda)-
 \frac{1}{4}\frac{1}{(H^2-\Lambda)^{3/2}} \Delta(\Lambda^2) + \frac{2}{V^2}
 \Delta(Vp_{\phi}) \,. \nonumber
\end{eqnarray}
(We use $-C_{\phi}$ in order to align forward motion of $\phi$ with forward
motion of $\tau$.)  The chain rule then gives the proper-time equations
\begin{eqnarray}
  \frac{{\rm d}V}{{\rm d}\tau} &=& \frac{{\rm d}V}{{\rm d}\phi} \frac{{\rm
      d}\phi}{{\rm d}\tau}  = 2VH+ 2 \Delta(VH)+
  2\frac{VH\Lambda}{(H^2-\Lambda)^2} \Delta(H^2) \\
 && - 2\frac{Hp_{\phi}}{V^2\sqrt{H^2-\Lambda}} \Delta(V^2) +
 2\frac{H}{V\sqrt{H^2-\Lambda}} \Delta(Vp_{\phi}) \\
 && -\frac{V(H^2+\Lambda)}{(H^2-\Lambda)^2} \Delta(H\Lambda)+ \frac{1}{2}
 \frac{VH}{(H^2-\Lambda)^2} \Delta(\Lambda^2)
\end{eqnarray}
and
\begin{eqnarray}
 \frac{{\rm d}H}{{\rm d}\tau} &=& -2(H^2-\Lambda) - 2\frac{H}{V} \Delta(VH)+
 \frac{2\Lambda}{H^2-\Lambda} \Delta(H^2)\\
 && + 2  \frac{p_{\phi}\sqrt{H^2-\Lambda}}{V^3} \Delta(V^2)-
 2\frac{\sqrt{H^2-\Lambda}}{V^2} \Delta(Vp_{\phi})\\
 && + \frac{1}{V}\Delta(V\Lambda)- 2\frac{H}{H^2-\Lambda} \Delta(H\Lambda)+
 \frac{1}{2} \frac{1}{H^2-\Lambda} \Delta(\Lambda^2)\,.
\end{eqnarray}

Alternatively, we could square the deparameterized quantum Hamiltonian
(\ref{Hphi}) and rearrange terms so as to make the expression look like the
classical constraint plus moment terms. We obtain
\begin{eqnarray} \label{HphiSquared}
 0 &=& \frac{H_{\phi}^2}{V}-V^2(H^2-\Lambda)-2H \Delta(VH)+
 \frac{V\Lambda}{H^2-\Lambda} \Delta(H^2)\\
&&+ \Delta(V\Lambda)- 
   \frac{VH}{H^2-\Lambda} \Delta(H\Lambda)+ \frac{1}{4}
   \frac{V}{H^2-\Lambda} \Delta(\Lambda^2)\,. \nonumber
\end{eqnarray}
It is then possible to treat $H_{\phi}=\langle\hat{p}_{\phi}\rangle$ as the
momentum of $\phi$ because, kinematically, $\{\phi,H_{\phi}\}= -i\hbar^{-1}
\langle[\hat{\phi},\hat{p}_{\phi}]\rangle=1$ in the effective framework. This
gives
\begin{eqnarray}
 \frac{{\rm d}\phi}{{\rm d}\tau} &=& -2\frac{H_{\phi}}{V} =
 2\sqrt{H^2-\Lambda}+ \frac{2H}{V\sqrt{H^2-\Lambda}}\Delta(VH)-
 \frac{\Lambda}{(H^2-\Lambda)^{3/2}} \Delta(H^2)\\
 && - \frac{1}{V\sqrt{H^2-\Lambda}} \Delta(V\Lambda)+
 \frac{H}{(H^2-\Lambda)^{3/2}} \Delta(H\Lambda)-
 \frac{1}{4}\frac{1}{(H^2-\Lambda)^{3/2}} \Delta(\Lambda^2)
\end{eqnarray}
and
\begin{eqnarray}
  \frac{{\rm d}V}{{\rm d}\tau} &=& 2VH+ 2 \Delta(VH)+
  2\frac{VH\Lambda}{(H^2-\Lambda)^2} \Delta(H^2) \\
 && -\frac{V(H^2+\Lambda)}{(H^2-\Lambda)^2} \Delta(H\Lambda)+ \frac{1}{2}
 \frac{VH}{(H^2-\Lambda)^2} \Delta(\Lambda^2)\,,\\
 \frac{{\rm d}H}{{\rm d}\tau} &=& -2(H^2-\Lambda) - 2\frac{H}{V} \Delta(VH)+
 \frac{2\Lambda}{H^2-\Lambda} \Delta(H^2)\\
 && + \frac{1}{V}\Delta(V\Lambda)- 2\frac{H}{H^2-\Lambda} \Delta(H\Lambda)+
 \frac{1}{2} \frac{1}{H^2-\Lambda} \Delta(\Lambda^2)\,.
\end{eqnarray}
These equations are different than what we obtained with the first choice of
$C$.

\subsection{Cosmological time}

For internal time $T$, we solve the constraint $C=0$ for the momentum
\begin{equation}
 \Lambda(V,H,p_{\phi}) = H^2-\frac{p_{\phi}^2}{V^2}\,.
\end{equation}
Its semiclassical quantization gives the Hamiltonian
\begin{equation} \label{HT}
 H_T = H^2-\frac{p_{\phi}^2}{V^2}+ \Delta(H^2)- \frac{3p_{\phi}^2}{V^4}
 \Delta(V^2)- \frac{1}{V^2} \Delta(p_{\phi}^2)+ 4\frac{p_{\phi}}{V^3}
 \Delta(Vp_{\phi})\,,
\end{equation}
generating equations of motion
\begin{equation}
 \frac{{\rm d}V}{{\rm d}T} = -2H
\end{equation}
and
\begin{equation}
 \frac{{\rm d}H}{{\rm d}T} = 2\frac{p_{\phi}^2}{V^3}+ 12\frac{p_{\phi}^2}{V^5}
 \Delta(V^2)+ \frac{2}{V^3}\Delta(p_{\phi}^2)- 12\frac{p_{\phi}}{V^4}
 \Delta(Vp_{\phi})\,.
\end{equation}

We attempt to transform to proper time using
\begin{equation}
 \frac{{\rm d}T}{{\rm d}\tau} = \{T,-C\}=-V\,.
\end{equation}
No quantum corrections appear in this equation because the constraint is
linear in $\Lambda$. We obtain
\begin{equation}
 \frac{{\rm d}V}{{\rm d}\tau} = 2VH
\end{equation}
and
\begin{eqnarray}
 \frac{{\rm d}H}{{\rm d}\tau} &=& -2\frac{p_{\phi}^2}{V^2}-
 12\frac{p_{\phi}^2}{V^4} \Delta(V^2)- \frac{2}{V^2}\Delta(p_{\phi}^2)+
 12\frac{p_{\phi}}{V^3} \Delta(Vp_{\phi})\\
&\approx& -2(H^2-\Lambda)- 2\Delta(H^2)- 6\frac{H^2-\Lambda}{V^2}\Delta(V^2)+
4\frac{p_{\phi}}{V^3} \Delta(Vp_{\phi})\,.
\end{eqnarray}

In the last step, we have used the constraint $H_T-\Lambda=0$ in order to
bring the equation closer to the form seen with $\phi$ as internal
time. Nevertheless, there is no obvious relationship between the two
deparameterizations (in either one of the two versions presented for the
scalar time), and covariance remains unclear.

\subsection{A new scalar time}

A formal difference between the scalar and cosmological choices of internal
times is the linear appearance of the time momentum in the former case,
compared with the quadratic appearence in the latter. In order to show that
this is not the reason for the disagreement of proper-time evolutions, we
modify the treatment of scalar time by applying a canonical transformation:
We replace $\phi$ and $p_{\phi}$ by $q:=\frac{1}{2}\phi/p_{\phi}$ and
$p:=p_{\phi}^2$. The constraint
\begin{equation}
 C=-VH^2+\frac{p}{V}+V\lambda=0
\end{equation}
is then linear in $p$ which we now use as the momentum of internal time $q$.

Proceeding as before, we have the quantum Hamiltonian
\begin{equation}
 H_q=V^2(H^2-\Lambda)+ (H^2-\Lambda)\Delta(V^2)+ 4VH\Delta(VH)+
 V^2\Delta(H^2)- 2V\Delta(V\Lambda)
\end{equation}
and the internal-time evolution equations
\begin{eqnarray}
 \frac{{\rm d}V}{{\rm d}q}&=& -2V^2H-2H\Delta(V^2)-4V\Delta(VH)\,,\\
 \frac{{\rm d}H}{{\rm d}q}&=&
 2V(H^2-\Lambda)+4H\Delta(VH)+2V\Delta(H^2)-2\Delta(V\Lambda)\,.
\end{eqnarray}
Internal time $q$ is tentatively related to proper time $\tau$ by
\begin{equation}
 \frac{{\rm d}q}{{\rm d}\tau} = -\frac{1}{V}\,,
\end{equation}
and we obtain proper-time equations
\begin{eqnarray}
 \frac{{\rm d}V}{{\rm d}\tau} &=& 2VH+2\frac{H}{V}\Delta(V^2)+4\Delta(VH)\,,\\
 \frac{{\rm d}H}{{\rm d}\tau} &=& -2(H^2-\Lambda)-4\frac{H}{V}
 \Delta(VH)-2\Delta(H^2)+ \frac{2}{V}\Delta(V\Lambda)
\end{eqnarray}
which agree with none of the previous versions.

\section{Gauge structure}
\label{s:Gauge}

Covariance is a property of the gauge nature of a theory. For systems with a
single Hamiltonian constraint $C$, as in our classical model,
reparameterization invariance is guaranteed by the fact that we always have
$\{C,C\}=0$ and the constraint is first class. It generates a gauge
transformation which corresponds to reparameterization invariance of the time
variable, be it proper time as the gauge parameter in ${\rm d}/{\rm
  d}\tau=\{\cdot,C\}$ or internal time. Even if the classical constraint is
modified by putative quantum corrections, as a single constraint it always
commutes with itself and reparameterization invariance should be
respected. Our examples contradict this expectation.

The discrepancy is resolved if we remember that quantization introduces new
degrees of freedom, parameterized in the effective formulation by
fluctuations, covariances and higher moments of a state. If fluctuations are
included as in our examples, the system is therefore equipped with a
different, enlarged phase space.

For the same reduction of degrees of freedom to result in this enlarged
setting as in the classical theory, there must also be additional
constraints. If a canonical pair such as $(\phi,p_{\phi})$ is eliminated by
solving the classical constraint and factoring out its gauge flow, not only
the expectation values of $\phi$ and $p_{\phi}$  must be eliminated by
quantized constraints but also the moments involving $\phi$ or $p_{\phi}$. On
the quantum phase space, these latter variables are independent of the
expectation values, and therefore require new constraints in order to be
eliminated.

\subsection{Effective constraints}

Using the canonical effective description, additional constraints appear
automatically for any first-class classical constraint $C$. If $\hat{C}$ is an
operator with classical limit $C$, about which we again assume only that it is
Weyl ordered, not only the expectation value 
\begin{equation}
 C_1:=\langle\hat{C}\rangle=0
\end{equation}
is a constraint, but also all expressions of the form
\begin{equation}
 C_f:=\langle(\hat{f}-f)\hat{C}\rangle=0
\end{equation}
where $f$ is an arbitrary classical phase-space function and $\hat{f}$ its
(Weyl-ordered) quantization. For $f\not=1$, the equation $C_f=0$ is
independent of $C_1=0$ on the quantum phase space. There are therefore
infinitely many new constraints $C_f$, which can conveniently be organized by
using for $f$ polynomials in some set of basic phase-space variables.

Just as expectation values of Hamiltonians used in the deparameterized
models, the effective constraints can be expanded in moments. We have
\begin{eqnarray}
 C_1(O_1, \ldots, O_n,\Delta(\cdot)) &=&
 C(O_1, \ldots, O_n)\\
&&+
 \sum_{a_1,\ldots,a_n} \frac{1}{a_1!\cdots a_n!}
 \frac{\partial^{a_1+\cdots+a_n} C(O_1, \ldots, O_n)}{\partial
   O_1^{a_1}\cdots \partial O_n^{a_n}}
 \Delta(O^{a_1}\cdots O^{a_n}) \nonumber
\end{eqnarray}
where the basic variables are called $O_1,\ldots,O_n$, $\Delta(\cdot)$ denotes
their moments, and $C$ is the classical constraint.
Similarly, any $C_f$ can be expanded in this way, but it usually requires
reordering terms because $\hat{f}\hat{C}$ is not necessarily Weyl ordered for
Weyl ordered $\hat{f}$ and $\hat{C}$. We will see this more explicitly in our
examples.

\subsection{Cosmological model}

We now compute effective constraints up to second-order moments for our
constraint (\ref{C}). This order requires us to accompany
$C_1=\langle\hat{C}\rangle$ by all constraints $C_f$ with $f$ linear in basic
variables. We obtain seven constraints
\begin{eqnarray}
 C_1 &=& -VH^2+\frac{p_{\phi}^2}{V}+V\Lambda
 +\frac{p_{\phi}^2}{V^3}\Delta(V^2)- 2H\Delta(VH)- V\Delta(H^2)\\
&&+
 \frac{1}{V}\Delta(p_{\phi}^2)- 2\frac{p_{\phi}}{V^2} \Delta(Vp_{\phi})+
 \Delta(V\Lambda)\,,\nonumber\\
 C_V &=& -\left(H^2+\frac{p_{\phi}^2}{V^2}-\Lambda\right) \Delta(V^2)- 2VH
 \left(\Delta(VH)-\frac{1}{2}i\hbar\right)\\
&&+
 2\frac{p_{\phi}}{V}\Delta(Vp_{\phi}) +V\Delta(V\Lambda)\,,\nonumber\\
 C_H &=& -2VH\Delta(H^2)- \left(H^2+\frac{p_{\phi}^2}{V^2}-\Lambda\right)
 \left(\Delta(VH)+\frac{1}{2}i\hbar\right)\\
&& +2\frac{p_{\phi}}{V}
 \Delta(Hp_{\phi})+ V\Delta(H\Lambda)\,,\nonumber\\
 C_{\phi} &=& -\left(H^2+\frac{p_{\phi}^2}{V^2}-\Lambda\right) \Delta(V\phi)-
 2VH\Delta(H\phi)\\
&&+ 2\frac{p_{\phi}}{V}\left(\Delta(\phi
   p_{\phi})+\frac{1}{2}i\hbar\right) +V\Delta(\phi\Lambda)\,,\nonumber\\
 C_{p_{\phi}} &=&
 -\left(H^2+\frac{p_{\phi}^2}{V^2}-\Lambda\right)\Delta(Vp_{\phi})
 -2VH\Delta(Hp_{\phi})+
 2\frac{p_{\phi}}{V}\Delta(p_{\phi})^2+V\Delta(p_{\phi}\Lambda)\,,\\
C_T &=& -\left(H^2+\frac{p_{\phi}^2}{V^2}-\Lambda\right)
\Delta(VT)-2VH\Delta(HT) +2\frac{p_{\phi}}{V}\Delta(p_{\phi}T)\\
&&+
V\left(\Delta(\Lambda T)+\frac{1}{2}i\hbar\right)\,,\nonumber\\
C_{\Lambda} &=& -\left(H^2+\frac{p_{\phi}^2}{V^2}-\Lambda\right)
\Delta(V\Lambda)- 2VH\Delta(H\lambda)+ 2\frac{p_{\phi}}{V}
\Delta(p_{\phi}\Lambda)+ V\Delta(\Lambda^2)\,.
\end{eqnarray}

The terms of $\frac{1}{2}i\hbar$ are from reordering to Weyl ordered
moments. Some of the effective constraints are therefore complex, and so will
be some of the moments after solving the constraints. This property is not
problematic because we have not eliminated any variables yet and are therefore
still in the kinematical setting. As shown in \cite{EffCons,EffConsRel}, after
solving the constraints and factoring out their gauge flows one can impose
reality conditions on the resulting physical moments. Real-valued observables
are then obtained, corresponding to expressions taken in the physical Hilbert
space.

Also in \cite{EffCons,EffConsRel}, it has been shown that the effective
constraints form a first-class system. Therefore, they generate gauge
transformations. However, the phase space of expectation values and moments up
to a certain order is not always symplectic, and the number of constraints is
not always equal to the number of independent gauge transformations. (See
\cite{brackets} for a discussion of first-class constraints in non-symplectic
systems.) In particular, a smaller number of gauge-fixing conditions may
be required if one would like to fix the gauge of a given set of constraints
on a Poisson manifold.

\subsection{Effective deparameterization}

Deparameterization with respect to a given internal time such as $\phi$
amounts to a specific choice of gauge fixing. After deparameterization,
$\phi$, just as the usual $t$ in non-relativistic quantum mechanics, is no
longer represented by an operator but only appears as a parameter in the
theory. It is not subject to quantum fluctuations and does not have quantum
correlations with other variables. These properties are reflected in the
gauge-fixing conditions
\begin{equation} \label{phiGauge}
 \Delta(\phi^2)=\Delta(V\phi)=\Delta(H\phi)= \Delta(\phi
 T)=\Delta(\phi\Lambda)=0
\end{equation}
which, as shown in \cite{EffTime,EffTimeLong}, suffice to fix the
effective constraints $C_f$ with linear $f$. 

The remaining covariance of $\phi$ with $p_{\phi}$ is not zero but takes the
complex value
\begin{equation} \label{Deltaphip}
 \Delta(\phi p_{\phi})=-\frac{1}{2}i\hbar
\end{equation}
as a consequence of $C_{\phi}=0$ together with the gauge-fixing
conditions. This complex value plays only a formal role, but it is useful
because it means that the uncertainty relation
\begin{equation}
 \Delta(\phi^2)\Delta(p_{\phi})^2-\Delta(\phi p_{\phi})^2\geq
 \frac{\hbar^2}{4}
\end{equation}
is still respected even with $\Delta(\phi^2)=0$.

\subsubsection{Scalar time}

We proceed to solving the remaining effective constraints. From $C_V=0$, we
obtain
\begin{equation} \label{DeltaVp}
 \Delta(Vp_{\phi}) = \frac{1}{2} \frac{V}{p_{\phi}}
 \left(\left(H^2+\frac{p_{\phi}^2}{V^2}-\Lambda\right) \Delta(V^2)+ 2VH
   \left(\Delta(VH)-\frac{1}{2}i\hbar\right)- V\Delta(V\Lambda)\right)\,;
\end{equation}
from $C_H=0$,
\begin{equation} \label{DeltaHp}
 \Delta(Hp_{\phi}) = \frac{1}{2}\frac{p_{\phi}}{V} \left(2VH\Delta(H^2)+
   \left(H^2+\frac{p_{\phi}^2}{V^2}-\Lambda\right)
   \left(\Delta(VH)+\frac{1}{2}i\hbar\right)- V\Delta(H\Lambda)\right)\,;
\end{equation}
from $C_{\Lambda}=0$,
\begin{equation}
 \Delta(p_{\phi}\Lambda) = \frac{1}{2}\frac{p_{\phi}}{V}
 \left(\left(H^2+\frac{p_{\phi}^2}{V^2}-\Lambda\right)\Delta(V\Lambda)+
   2VH\Delta(H\Lambda)- V\Delta(\Lambda^2)\right)\,;
\end{equation}
and from $C_{p_{\phi}}=0$,
\begin{eqnarray}
 \Delta(p_{\phi}^2) &=& \frac{1}{2} \frac{p_{\phi}}{V}
 \left(\left(H^2+\frac{p_{\phi}^2}{V^2}-\Lambda\right)
   \Delta(Vp_{\phi})+2VH\Delta(Hp_{\phi}) - V\Delta(p_{\phi}\Lambda)\right)\\
 &=& \frac{1}{4} \frac{V^2}{p_{\phi}^2}
 \left(H^2+\frac{p_{\phi}^2}{V^2}-\Lambda\right) \Delta(V^2)+
 \frac{V^4H^2}{p_{\phi}^2} \Delta(H^2)+ \frac{V^3H}{p_{\phi}^2}
 \left(H^2+\frac{p_{\phi}^2}{V^2}-\Lambda\right)\Delta(VH)\nonumber\\
 &&-
 \frac{V^3}{p_{\phi}^2} \left(H^2+\frac{p_{\phi}^2}{V^2}-\Lambda\right)
 \Delta(V\Lambda)- \frac{V^4H}{p_{\phi}^2} \Delta(H\Lambda)+ \frac{1}{4}
 \frac{V^4}{p_{\phi}^2} \Delta(\Lambda^2)\,.
\end{eqnarray}

Notice again that the moments $\Delta(Vp_{\phi})$ and $\Delta(Hp_{\phi})$ are
complex. The reason is that we are in the process of deparameterizing by
$\phi$, which eliminates all moments related to the canonical pair
$(\phi,p_{\phi})$, including their covariances with other variables. In the
complex moments, $p_{\phi}$ is therefore not an independent variable
anymore. It is a function of $V$, $H$, $\Lambda$ and the moments owing to the
constraint $C_1=0$. While $\hat{V}\hat{p}_{\phi}$ is a Hermitian operator
when $V$ and $p_{\phi}$ are independent, it is no longer Hermitian in this
ordering if $p_{\phi}$ is a function of $H$ after solving $C_1=0$. The complex
contributions to $\Delta(Vp_{\phi})$ and $\Delta(Hp_{\phi})$ implicitly
describe the ordering obtained after solving the constraints. Note that
$\Delta(p_{\phi}\Lambda)$ remains real, which is consistent with the fact that
$p_{\phi}$ does not depend on $T$ after $C_1=0$ is solved. (See also
\cite{FluctEn} for a related discussion of complex moments.)

All $p_{\phi}$-moments can now be eliminated from the remaining constraint
$C_1=0$, as appropriate for a system deparameterized with respect to
$\phi$. The resulting expression can be compared with the evolution generator
on the physical Hilbert space, where no operators for $\phi$ and $p_{\phi}$
exist.  However, there is one last step before such a comparison can be
done. We have introduced gauge-fixing conditions, and must therefore make sure
that the evolution generator preserves these conditions. Usually, such a
generator is not the remaining (unfixed) constraint $C_1$ but a linear
combination of all the constraints of the system. (The gauge fixing requires
us to use a specific lapse function $N$ on the quantum phase space.)

Using the methods of \cite{EffTimeLong,AlgebraicTime}, one can check that, in
the present example, the unique generator respecting the gauge-fixing
conditions is of the form
\begin{eqnarray} \label{NC}
  NC &=& \frac{1}{2p_{\phi}} \left((VC)_1- \frac{1}{2p_{\phi}} (VC)_{p_{\phi}}-
    \frac{1}{2p_{\phi}} \frac{\partial p_{\phi}}{\partial V} (VC)_V -
    \frac{1}{2p_{\phi}} \frac{\partial p_{\phi}}{\partial H} (VC)_H\right.\\
  &&\left.-
    \frac{1}{2p_{\phi}} \frac{\partial p_{\phi}}{\partial \Lambda}
    (VC)_{\Lambda} - 
    \frac{1}{2p_{\phi}} \frac{\partial p_{\phi}}{\partial T} (VC)_T\right)
  \nonumber 
\end{eqnarray}
where $(VC)_f$ are defined just like the previous effective constraints but
with $\hat{V}\hat{C}$ inserted instead of $\hat{C}$. The factor of $\hat{V}$
removes the $1/V$ in the quadratic kinetic term $p_{\phi}^2/V$ in
$C$. We emphasize that we are still dealing with the original system of
effective constraints because any $(VC)_f$ can be written as a linear
combination of the $C_f$ to the same order. For instance, 
\begin{equation}
 (VC)_1 = \langle(V+(\hat{V}-V))\hat{C}\rangle = VC_1+C_V
\end{equation}
and
\begin{equation}
 (VC)_V = \langle(\hat{V}-V)(V+(\hat{V}-V))\hat{C}\rangle = VC_V+\Delta(V^2)
 C_1\,. 
\end{equation}

For our present purposes, it suffices to justify the combination (\ref{NC}) of
constraints by confirming that the resulting generator
\begin{eqnarray}
 NC&=&\frac{p_{\phi}-V\sqrt{H^2-\Lambda}}{2p_{\phi}}
 \left(p_{\phi}+V\sqrt{H^2-\Lambda}\right) + \frac{H}{\sqrt{H^2-\Lambda}}
 \Delta(VH)-
 \frac{1}{2} \frac{V\Lambda}{(H^2-\Lambda)^{3/2}} \Delta(H^2) \nonumber\\
&&-
 \frac{1}{2\sqrt{H^2-\Lambda}} \Delta(V\Lambda)+ \frac{1}{2}
   \frac{VH}{(H^2-\Lambda)^{3/2}} \Delta(H\Lambda)- \frac{1}{8}
   \frac{V}{(H^2-\Lambda)^{3/2}} \Delta(\Lambda^2) 
\end{eqnarray}
indeed preserves the gauge-fixing conditions: all $p_{\phi}$-moments have
cancelled out.  Moreover, solving $NC=0$ for $p_{\phi}$ gives an expression
identical with the deparameterized $\phi$-Hamiltonian (\ref{Hphi}). We
therefore confirm that deparemeterization can be performed before or after
quantization, with equivalent results.

\subsubsection{Cosmological time}

Deparameterization of the effective constraints with respect to $T$ is done by
using the gauge-fixing conditions
\begin{equation} \label{TGauge}
 \Delta(T^2)=\Delta(VT)=\Delta(HT)=\Delta(\phi T)=\Delta(p_{\phi}T)=0
\end{equation}
which implies $\Delta(T\Lambda)=-\frac{1}{2}i\hbar$ using $C_T=0$. As before,
we can solve all constraints for the $\Lambda$-moments, but we do not need the
explicit expressions because the relevant generator,
\begin{equation}
 \left(V^{-1}C'\right)_1 = -H^2+\frac{p_{\phi}^2}{V^2}+\Lambda-
 \Delta(H^2)+3\frac{p_{\phi}^2}{V^4} \Delta(V^2)- 4\frac{p_{\phi}}{V^3}
 \Delta(Vp_{\phi})+ \frac{1}{V^2} \Delta(p_{\phi}^2)
\end{equation}
contains no such moments.  Solving $(V^{-1}C')_1=0$ for $\Lambda=H_T$ gives an
expression for the $T$-Hamiltonian identical with (\ref{HT}).

Similarly to the scalar case, the momentum $\Lambda$
appears with a factor of $V$, which leads to the modified effective constraint
$(V^{-1}C')_1$. We have indicated by the prime on $C'$ a change of factor
ordering with respect to the original Weyl-ordered constraint operator
$\hat{C}$. In order for $(V^{-1}C')_1$ to be real, we need a symmetric ordering
of the contribution $\hat{V}^{-1} (\hat{V}\hat{H}^2)'$ with some ordering of
$\hat{V}\hat{H}^2$ again indicated by the prime. The product with
$\hat{V}^{-1}$ is not symmetric if Weyl-ordering is used for
$(\hat{V}\hat{H}^2)'$, but it is symmetric if we instead use
\begin{equation} \label{VH2}
 \hat{V}\hat{H}^2 =
 \frac{1}{3}(\hat{V}\hat{H}^2+\hat{H}\hat{V}\hat{H}+\hat{H}^2\hat{V})-i\hbar
 \hat{H}= (\hat{V}\hat{H}^2)_{\rm Weyl}-i\hbar\hat{H}\,.
\end{equation}
Indeed, with the subtraction of $i\hbar \hat{H}$ in the reordered constraint
$\hat{C}'=\hat{C}-i\hbar\hat{H}$, we have 
\begin{equation}
 \left(V^{-1}C'\right)_1 =
 \langle(V^{-1}-V^{-2}(\hat{V}-V))(\hat{C}-i\hbar\hat{H})\rangle=
 \frac{C_1}{V}-\frac{C_V-i\hbar VH}{V^2}
\end{equation}
as a real expression of the effective constraints, where $C_V$ has imaginary
part $\hbar VH$.

Unlike the generator of deparameterized evolution in the scalar model, the
generator for cosmological time is {\em not} a linear combination of the
original effective constraints because $i\hbar H/V$ is not of such a form. The
two deparameterized models are therefore realized within the same effective
constrained system only if we ignore reordering contributions with an explicit
dependence on $\hbar$. The moment corrections in the two models are related
by a gauge transformation and therefore provide the same effects in
observables. However, $\hbar$-dependent terms are not related by gauge
transformations and lead to different effects in observables. For
semiclassical states, for which our analysis is valid, second-order moments
are generically of the order $\hbar$, and it is not possible to ignore factor
ordering corrections compared with moment corrections. The two different
internal times therefore lead to different predictions, and time
reparameterization invariance is broken.

\subsection{Proper time}

Using effective constraints, we have rederived the deparameterized
Hamiltonians (\ref{Hphi}) and (\ref{HT}) for our model with two different
choices of internal time. The agreement with derivations in deparameterized
models in the preceding Sec.~\ref{s:Deparam} demonstrates that it does not
matter whether we deparameterize the classical theory and then quantize the
internal-time Hamiltonians, or whether we quantize first using effective
constraints and then deparameterize. At least at the semiclassical level used
here, deparameterization therefore commutes with quantization.

Moreover, we have realized the two internal-time models as two different gauge
fixings of the same constrained system, up to reordering terms. Since the
constraints are first class, the observable content of the models does not
depend on the particular gauge fixing used to derive it, as long as only
moment corrections are considered. (Explicit gauge transformations of moments
relating the models can be derived as in \cite{EffTimeLong}.) We have
therefore demonstrated in our quantized cosmological model how covariance can
in principle be realized, in the sense that the two internal-time versions
derived in Sec.~\ref{s:Deparam} would be equivalent to each other. However, in
our explicit example, covariance is broken by factor ordering corrections,
which appear whenever the momenta of two internal times appear in the
constraint with different phase-space dependent factors. However, this result,
which we consider to be rather important, cannot explain the mismatch of
proper-time evolutions we found in Sec.~\ref{s:Deparam} because this mismatch
appears even for moment corrections. The existence of gauge transformations
that successfully transform the moment corrections in deparameterized
effective constraints, at first sight, makes the disagreement of their
proper-time evolutions only more puzzling.

However, supplied with the methods of effective constraints, we can now
revisit this question with a complete view on the gauge structure. Our first
attempt to derive proper-time evolution from internal-time evolution required
an expression for ${\rm d}\phi/{\rm d}\tau$ or ${\rm d}T/{\rm d}\tau$. Since
there is no $\tau$ in the deparameterized theory, such an expression can only
come from the original constraint. It may be amended by different versions of
moment corrections, as seen in the scalar example, but it is always closely
related to the original gauge generator which we have now called $C_1$. 

At this point, we can see the reason for our problem of mismatched proper-time
evolutions. A deparameterized model is equivalent to a specific gauge fixing
of effective constraints. The gauge fixing must be preserved by evolution in
the model, which requires a specific combination of effective constraints as
evolution generator. If the classical constraint is not linear in the momentum
of internal time, or if there are phase-space dependent factors such as $V$ or
$1/V$ of the momentum of internal time, the evolution generator preserving the
gauge fixing is not equal to the effective constraint $C_1$ used for proper
time. The only generator consistent with the gauge-fixing conditions is the
deparameterized Hamiltonian (or this Hamiltonian multiplied with a quantum
phase-space function not depending on internal time and its momentum). 

In this way, only the deparameterized evolution can be described within a
deparameterized model. It is impossible to transform this evolution to proper
time and still have reparameterization invariance or covariance. Referring to
the chain rule in order to transform from an internal time to proper time is
meaningless in this context of multiple constraints. The 1-parameter chain
rule ${\rm d}/{\rm d}\tau=({\rm d}\phi/{\rm d}\tau) {\rm d}/{\rm d}\phi$ is
valid only if evolution is described by a unique 1-dimensional
trajectory. This is the case in the classical theory, in which there is just
one constraint, but not in the quantum theory in which expectation values and
moments provide independent constraints. In order to apply the 1-parameter
chain rule, one would first have to select a unique trajectory generated by a
distinguished linear combination of the constraints. But once a specific
linear combination has been selected, it corresponds to a fixed choice of
time. Transformations between different time choices are then no longer
possible.

There is a way to obtain proper-time evolution from the effective
constraints. Proper time is not a phase-space variable, and therefore it does
not correspond to a natural gauge fixing of the effective constraints. 
Instead of fixing the gauge of linear constraints $C_f$, we compute invariant
expectation values and moments, or Dirac observables of this subset of
constraints. Up to terms of higher order in $\hbar$ including products of
second-order moments, as always in this paper, we have the invariants
\begin{eqnarray}
 {\cal V} &=& V-\frac{VH}{p_{\phi}} \Delta(V\phi)-
 \frac{V^2}{p_{\phi}}\Delta(H\phi)+
 \frac{V^3\Lambda}{2p_{\phi}^2}\Delta(\phi^2)\,,\\
 {\cal H} &=& H+2\frac{VH}{p_{\phi}} \Delta(H\phi)- \frac{V}{p_{\phi}}
 \Delta(\phi\Lambda)+ H\Delta(\phi^2)
\end{eqnarray}
as well as
\begin{eqnarray}
 \Delta({\cal V}^2) &=& \Delta(V^2)-2\frac{V^2H}{p_{\phi}} \Delta(V\phi)+
 \frac{V^4H^2}{p_{\phi}^2} \Delta(\phi^2)\,,\\
 \Delta({\cal VH}) &=& \Delta(VH)+\frac{p_{\phi}}{V} \Delta(V\phi)-
 \frac{V^2H}{p_{\phi}} \Delta(H\phi)- HV\Delta(\phi^2)\,,\\
 \Delta({\cal H}^2) &=& \Delta(H^2)+2\frac{p_{\phi}}{V} \Delta(H\phi)+
 \frac{p_{\phi}^2}{V^2} \Delta(\phi^2)\,,\\
 \Delta({\cal V}p_{\phi}) &=& \Delta(Vp_{\phi})-
 \frac{V^2H}{p_{\phi}} \Delta(\phi p_{\phi})\,. \label{DeltaVpObs}
\end{eqnarray}
Moreover, $p_{\phi}$, $\Lambda$, $\Delta(p_{\phi}^2)$,
$\Delta(p_{\phi}\Lambda)$ and $\Delta(\Lambda^2)$ are invariant. Note that
$\Delta({\cal V}p_{\phi})$ in (\ref{DeltaVpObs}) is real even if $\phi$ is
used as internal time because the non-zero imaginary parts of
$\Delta(Vp_{\phi})$ and $\Delta(\phi p_{\phi})$, according to (\ref{DeltaVp})
and (\ref{Deltaphip}) cancel out completely.

These combinations of expectation values and moments are invariant to
second-order moments under gauge transformations generated by effective
constraints $C_f$ with $f$ linear in basic variables. In the gauge
(\ref{phiGauge}) of a formulation deparameterized by internal time $\phi$,
they are equal to the kinematical expectation values and moments of the same
type and thus provide an invariant extension of these variables. In the gauge
of some other internal time such as $T$, with conditions (\ref{TGauge}), there
are additional non-zero moments compared with the simple kinematical
expressions $V$, $H$, $\Delta(V^2)$, $\Delta(VH)$, $\Delta(H^2)$ and
$\Delta(Vp_{\phi})$. If one analyzes a model using different internal times,
such as $\phi$ and $T$ in the present case, one should therefore not directly
compare moments of the same type, but combinations as dictated by invariant
moments. For instance, the fluctuation $\Delta(V^2)$ computed with internal
time $\phi$ represents the same observable (with respect to linear constraints
$C_f$) as $\Delta(V^2)-2(V^2H/p_{\phi}) \Delta(V\phi)+ (V^4H^2/p_{\phi}^2)
\Delta(\phi^2)$ computed with internal time $T$.

The remaining constraint $C_1$ written in terms of invariant expectation
values and moments is
\begin{equation}
 {\cal C} = -{\cal V}{\cal H}^2+ \frac{p_{\phi}^2}{{\cal V}^2}+ {\cal
   V}\Lambda- {\cal V}\Delta({\cal H}^2)- 2{\cal H}\Delta({\cal VH})+
 \frac{p_{\phi}^2}{{\cal V}^3} \Delta({\cal V}^2) +\frac{1}{{\cal V}}
 \Delta(p_{\phi}^2) -2\frac{p_{\phi}}{{\cal V}^2} \Delta({\cal V}p_{\phi})+
 \Delta({\cal V}\Lambda) + i\hbar H\,.
\end{equation}
The moment corrections are of the same form that $C_1$ has in terms of the
kinematical expectation values and moments. However, the transformation to
invariant moments leads to an imaginary part $\hbar H$ which indicates that
the Weyl-ordered operator used for $C_1$ was not of the correct ordering.
Similarly to (\ref{VH2}), we have
\begin{equation} \label{H2V}
 \hat{H}^2\hat{V} =
 \frac{1}{3}(\hat{V}\hat{H}^2+\hat{H}\hat{V}\hat{H}+\hat{H}^2\hat{V})+i\hbar
 \hat{H}= (\hat{V}\hat{H}^2)_{\rm Weyl}+i\hbar\hat{H}\,.
\end{equation}
If we use the ordering $(\hat{V}\hat{H}^2)''=\hat{H}^2\hat{V}$ in a reordered
constraint operator $\hat{C}''$, we have $\hat{C}''=\hat{C}_{\rm
  Weyl}-i\hbar\hat{H}$ and the imaginary parts in $\langle\hat{C}''\rangle$
cancel out after transformation to invariant expectation values and moments:
\begin{equation}
 {\cal C}'' = -{\cal V}{\cal H}^2+ \frac{p_{\phi}^2}{{\cal V}^2}+ {\cal
   V}\Lambda- {\cal V}\Delta({\cal H}^2)- 2{\cal H}\Delta({\cal VH})+
 \frac{p_{\phi}^2}{{\cal V}^3} \Delta({\cal V}^2) +\frac{1}{{\cal V}}
 \Delta(p_{\phi}^2) -2\frac{p_{\phi}}{{\cal V}^2} \Delta({\cal V}p_{\phi})+
 \Delta({\cal V}\Lambda)\,.
\end{equation}
We have not found an independent argument why the ordering of $\hat{C}''$
should be used for proper-time evolution. The appearance of this particular
ordering is therefore rather surprising, as is the fact that it is different
from the two orderings required for scalar and cosmological internal times.

With a real-valued effective constraint ${\cal C}''$ in terms of invariants,
we can finally introduce proper-time evolution.  We do not introduce
gauge-fixing conditions but explicitly select the lapse function of the
generic evolution generator
\begin{equation}
  NC_{\rm eff}=N_1C_1''+ \sum_f N_fC_f''= \langle (N_1 +N_f
  (\hat{f}-f))\hat{C}''\rangle =\langle\hat{N}\hat{C}''\rangle
\end{equation}
by setting all $N_f=0$ for $f\not=1$ and $N_1=1$. This choice
implements the feature that proper time, in a geometrical formulation,
corresponds to a lapse function $N=1$. At the operator level, we
should then have $\hat{N}=1$ without any contributions from
$\hat{f}-f$. Proper-time evolution equations are then generated by
${\cal C}''$, which is $\langle\hat{C}''\rangle$ expressed in
invariant expectation values and moments. Just as in classical
equations, it is not necessary to compute complete Dirac observables
which are also invariant under the flow generated by ${\cal C}''$,
since we can directly interpret proper-time trajectories in
geometrical terms. The tedious constructions of physical Hilbert
spaces in standard treatments of canonical quantum cosmology are, at
the effective level, replaced by invariance conditions with respect to
the flow generated by $C_f$, combined with reality conditions on
${\cal C}''$.

Proper time can therefore be implemented within the effective
constrained system, but it amounts to a gauge fixing different from
most deparameterized models. If we consider only moment corrections,
there are gauge transformations between proper-time and all
deparameterized models within the effective constrained system and
reparameterization invariance is preserved, including proper
time. Factor-ordering corrections generically break
covariance. However, no gauge transformation to proper time exists
within a deparameterized model, in which the gauge fixing can no
longer be changed. This is the case even if factor ordering terms are
ignored, so that covariance is more strongly broken in such cases.

Other coordinate times, such as conformal time, can be implemented in the same
way by still using $N_f=0$ but $N\not=1$ a function of expectation
values. Their evolution generators are given by ${\cal N}{\cal C}''$ where
${\cal N}$ is obtained by replacing expectation values in $N$ by their
invariant analogs. No new factor ordering of $\hat{C}$ is required because we
just multiply the proper-time generator ${\cal C}''$ with a function of
invariants, which keeps the expression real. Our definition of proper-time
therefore allows the same changes of time coordinates as in the classical
theory and is, in this sense, time reparameterization invariant. This
invariance is broken only if we try to compare coordinate time with internal
time.

\section{Discussion}

We have pointed out that time reparameterization invariance of
effective equations is not guaranteed after quantization even in
systems with a single constraint, and illustrated this often
overlooked property in a specific cosmological model. Our detailed
analysis of the underlying quantum gauge system has led us to a new
procedure in which one can implement proper-time evolution at the
effective level. This new definition includes all analogs of different
classical choices of coordinate time and is time reparameterization
invariant in this sense. Moreover, our procedure unifies models with
coordinate times and internal times because they are all obtained from
the same first-class constrained system by imposing different gauge
conditions, up to factor orderings.

The last condition is important and ultimately leads to violations of
time reparameterization invariance or covariance of internal-time
formulations. The effective constrained system provides gauge
transformations that map moment corrections in an evolution generator
for one time choice to the moment corrections obtained with a
different time choice, including proper time. However, in our model,
the time choices we studied explicitly, given by scalar time,
cosmological time and proper time, all require different factor
orderings of the constraint operator for real evolution
generators. Since effective constraints are computed for a given
factor ordering of the constraint operator, they do not allow gauge
transformations that would change factor ordering corrections. Factor
ordering terms therefore generically imply that different time choices
lead to different predictions, and time reparameterization invariance
of internal-time formulations is broken. The only solution to this
important problem is to insist on one specific time choice for all
derivations. The only distinguished time choice, in our opinion, is
proper time: it refers directly to the time experienced by observers
and gives evolution equations that can be used directly in an
effective Friedmann equation of cosmological models. Moreover, it is
time-reparameterization invariant when compared with other choices of
coordinate time, while there are no complete transformations for
different choices of internal time.

We have worked entirely at an effective level up to second order in moment
corrections, corresponding to a semiclassical approximation to first order in
$\hbar$. This order suffices to demonstrate our claims because differences in
quantum corrections between the models are visible at this order. In
principle, one can extend the effective expansion to higher orders, but it
becomes more involved and is then best done using computational help. We have
not considered such an extension in the present paper because the orders we
did include already show quite dramatic differences between the models if
improper gauge conditions are used, for instance by trying to rewrite a
deparameterized model in proper time by using the 1-parameter chain rule.

Our deparameterized models could certainly be formulated with operators acting
on a physical Hilbert space without using an effective theory. However, no
general method is known that would allow one to compare physical Hilbert
spaces based on different deparameterizations, or to introduce proper time at
this level. By using an effective formulation, we have gained the advantage of
being able to embed all such models within the same constrained system, and to
transform their moment corrections by simple changes of gauge
conditions. These properties were crucial in our strict definition of
proper-time evolution at the quantum level, for which we used effective
observables such as invariant moments instead of operators on a physical
Hilbert space. Internal-time formulations based on a single physical Hilbert
space, as used for instance in loop quantum cosmology, cannot be assumed to
give correct moment terms in effective equations, strengthening the
results of \cite{MultChoice}. Investigations of internal-time formulations of
quantum cosmological models with significant quantum fluctuations are
therefore likely to be spurious.

\section*{Acknowledgements}

This work was supported in part by NSF grant PHY-1607414 and a McNair
scholarship.


\end{document}